\documentclass[11pt]{article}
\usepackage{latexsym}
\usepackage{graphicx}
\usepackage{psfrag}
\newcommand{\be}{\begin{equation}}
\newcommand{\ee}{\end{equation}\noindent}
\newcommand{\bear}{\begin{eqnarray}}
\newcommand{\ear}{\end{eqnarray}\noindent}

\date{}
\renewcommand{\theequation}{\arabic{equation}}

\def\Eins{\mathord{1\hskip -1.5pt
\vrule width .5pt height 7.75pt depth -.2pt \hskip -1.2pt
\vrule width 2.5pt height .3pt depth -.05pt \hskip 1.5pt}}

\newcommand{\slD}{\raise.15ex\hbox{$/$}\kern-.57em\hbox{$D$}}
\newcommand{\slpartial}{\raise.15ex\hbox{$/$}\kern-.57em\hbox{$\partial$}}
\newcommand{\slG}{{{\dot G}\!\!\!\! \raise.15ex\hbox {/}}}

\def\non{\nonumber}
\def\beqn*{\begin{eqnarray*}}
\def\eqn*{\end{eqnarray*}}

\def\square{\kern1pt\vbox{\hrule height 1.2pt\hbox{\vrule width 1.2pt
   \hskip 3pt\vbox{\vskip 6pt}\hskip 3pt\vrule width 0.6pt}
   \hrule height 0.6pt}\kern1pt}
\def\slash#1{#1\!\!\!\raise.15ex\hbox {/}}
\def\slashleft#1{#1\!\!\!\!\raise.15ex\hbox {/}}

\def\dps{\displaystyle}

\def\half{{1\over 2}}

\def\e{\mbox{e}}

\def\4piTD{{(4\pi T)}^{-{D\over 2}}}
\def\4piT4{{(4\pi T)}^{-2}}

\def\Tintm4{{\dps\int_{0}^{\infty}}{dT\over T}\,e^{-m^2T}
    {(4\pi T)}^{-2}}
\def\Tintm{{\dps\int_{0}^{\infty}}{dT\over T}\,e^{-m^2T}}

\def\be{\begin{equation}}\def\ee{\end{equation}}
\def\bea{\begin{eqnarray}}\def\eea{\end{eqnarray}}
\def\ba{\begin{array}}\def\ea{\end{array}}
\def\bea{\begin{eqnarray}}\def\barr{\begin{array}}\def\earr{\end{array}}
\def\eea{\end{eqnarray}}
\begin{document} 
\newcommand{\ho}[1]{$\, ^{#1}$}
\newcommand{\hoch}[1]{$\, ^{#1}$}
\pagestyle{empty}
\renewcommand{\thefootnote}{\fnsymbol{footnote}}
\hfill {\sl BUCMP/02-04}

\hfill {\sl UMSNH-Phys/02-7}   
\vskip .4cm
\begin{center}
{\Large\bf Two-loop self-dual
Euler-Heisenberg Lagrangians (II):}\\ ~\\
{\Large\bf Imaginary part and Borel analysis}
\vskip1.3cm
{\large Gerald V. Dunne}
\\[1.5ex]
{\it
Department of Physics\\
University of Connecticut\\
Storrs CT 06269, USA
}
\vspace{.8cm}

 {\large Christian Schubert
}
\\[1.5ex]
{\it
Instituto de F\'{\i}sica y Matem\'aticas
\\
Universidad Michoacana de San Nicol\'as de Hidalgo\\
Apdo. Postal 2-82\\
C.P. 58040, Morelia, Michoac\'an, M\'exico\\
schubert@itzel.ifm.umich.mx\\
}
\vspace{.2cm}
{\it
Center for Mathematical Physics, Mathematics Department\\
Boston University, Boston, MA 02215, USA}\\
\vspace{.2cm}
{\it
California Institute for Physics and Astrophysics\\
366 Cambridge Ave., Palo Alto, CA 94306, USA}

\vskip 2cm
 %
 {\large \bf Abstract}
\end{center}
\begin{quotation}
\noindent

We analyze the structure of the imaginary part of the two-loop 
Euler-Heisenberg QED effective Lagrangian for a constant self-dual background. The
novel feature of the two-loop result, compared to one-loop, is that the prefactor of
each exponential (instanton) term in the imaginary part has itself an asymptotic
expansion.  We also perform a high-precision test of Borel summation techniques
applied to the weak-field expansion, and find that the Borel dispersion relations
reproduce the full prefactor of the leading imaginary contribution.

\end{quotation}
\vskip 1cm
\clearpage
\renewcommand{\thefootnote}{\protect\arabic{footnote}}
\pagestyle{plain}

\section{Introduction: imaginary part of effective Lagrangians}
\label{introduction}
\renewcommand{\theequation}{1.\arabic{equation}}
\setcounter{equation}{0}

The one-loop Euler-Heisenberg effective Lagrangians, for spinor QED
\cite{eulhei,weisskopf,schwinger51} and scalar QED \cite{schwinger51}, in a
constant background electromagnetic field, have the following well-known integral
representations:
\bear
{\cal L}_{\rm spin}^{(1)} (a,b)
&=&
-
{1\over 8\pi^2}
\int_0^{\infty}{dT\over T}
\,\e^{-m^2T}
\biggl[
{e^2ab\over \tanh(eaT)\tan(ebT)} -{1\over T^2}
-{e^2\over 3} (a^2-b^2)
\biggr]
\label{L1spinren}
\ear
\bear
{\cal L}_{\rm scal}^{(1)}(a,b) 
&=&
{1\over 16\pi^2}
\int_0^{\infty}{dT\over T}
\,\e^{-m^2T}
\biggl[
{e^2ab\over \sinh(eaT)\sin(ebT)}-{1\over T^2}
+{e^2\over 6} (a^2-b^2) 
\biggr]
\label{L1scalren}
\ear
Here $T$ denotes the (Euclidean) proper-time of the
loop fermion or scalar, and $a,b$ are related to the two
invariants of the Maxwell field by
$a^2-b^2=B^2-E^2, ab = {\bf E}\cdot{\bf B}$.
These Lagrangians have been renormalized \cite{eulhei,weisskopf,schwinger51} on-shell
by subtracting, under the integral, the terms of zeroeth and second order in $a,b$.

The effective Lagrangians (\ref{L1spinren}) and (\ref{L1scalren}) are
real for a purely magnetic field, while in the presence of an electric field there is
an imaginary (absorptive) part, indicating the process of electron--positron (resp.
scalar--antiscalar) pair creation by the field. The vacuum
persistence amplitude is $e^{i{\cal L}}$, so the probability of pair
production is approximately $2\,{\rm Im}\,{\cal L}$ \cite{schwinger51}. 
For example, in the case of a purely electric field, $E$, the effective Lagrangians 
(\ref{L1spinren}) and (\ref{L1scalren})  have imaginary part given by:
\begin{eqnarray}
{\rm Im} {\cal L}_{\rm spin}^{(1)}(E) &=&  \frac{m^4}{8\pi^3}
\beta^2\, \sum_{k=1}^\infty \frac{1}{k^2}
\,\exp\left[-\frac{\pi k}{\beta}\right]
\label{fullimag}\\
{\rm Im}{\cal L}_{\rm scal}^{(1)}(E) 
&=&
-\frac{m^4}{16\pi^3}
\beta^2\, \sum_{k=1}^\infty \frac{(-1)^{k}}{k^2}
\,\exp\left[-\frac{\pi k}{\beta}\right]
\label{L1scalim}
\end{eqnarray}
where $\beta = eE/m^2$. These expressions are clearly non-perturbative in terms of
the field and coupling.

The physical interpretation of these imaginary parts (\ref{fullimag}) and
(\ref{L1scalim}) of the effective Lagrangians is that the coefficient of the $k$-th
exponential can be directly identified with the rate for the coherent production of
$k$ pairs by the field \cite{schwinger51,nikishov}. This physical meaning of the
individual terms of these series follows most directly from the alternative
representation due to Nikishov \cite{nikishov,page},
\begin{eqnarray}
 VT\,{2\over \hbar}\, {\rm Im}\, {\cal L}^{(1)}(E)
 &=&
\mp \sum_r \int {V\, d^3p \over {(2\pi\hbar)}^3}
{\rm ln}(1\mp\bar n_p),\nonumber\\
\bar n_p &=& \exp \Bigl(-\pi {m^2 + p_{\perp}^2\over eE}\Bigr)
\label{repnikishov}
\end{eqnarray} 
Here $\bar n_p$ is the mean number of pairs produced by the field
in the state with given momentum $p$ and spin projection $r$, the
$\mp$ refers to the spinor/scalar cases respectively, and $p_\perp$ is the
momentum transverse to the field. An expansion of the logarithm in
$\bar n_p$ and term-by-term integration leads back to Schwinger's
formulas (\ref{fullimag}) and (\ref{L1scalim}). Thus the leading term
in this expansion can be interpreted as the mean number $\bar n_p$
of pairs in the unit 4-volume $VT$, while the higher ($k\ge 2$)
terms describe the coherent creation of $k$ pairs. This also explains the
physical origin of the alternating signs in the scalar case (\ref{L1scalim}).

The exponential suppression factors in (\ref{fullimag})
and (\ref{L1scalim}) are very small for the field strengths which are
presently possible for a macroscopic field in the laboratory. 
While positron production has recently been observed in an experiment
involving electrons traversing the focus of a terawatt laser
\cite{burkeetal}, in this experiment the electric field strength is such
that it is on the border-line between the perturbative and
nonperturbative regimes, and so the positron production can be explained
by multiphoton Compton scattering \cite{burkeetal}, rather than in terms
of nonperturbative pair creation \cite{mel}. This phenomenon has been
analyzed in detail in \cite{ringwald}, along with the prospects of
observing nonperturbative pair creation using an X-ray free-electron
laser which produces higher fields than those obtained by an optical
laser. We mention also a recent  analysis \cite{alkofer} which
considers nonequilibrium and back-reaction effects.

From a computational perspective, the imaginary parts of the effective
Lagrangian can be computed in several ways. At one-loop, the most direct
way is to consider the analytic properties of the integral
representations in (\ref{L1spinren}) and (\ref{L1scalren}). However, this
approach is quite diffcult at the two-loop level for an electric field
background, because the corresponding integral representations are double
integrals, with much more complicated integrands (see discussion below). 
An alternative approach is to exploit the well-known relation between the
large-order divergence of perturbation theory and non-perturbative
physics \cite{arkady,leguillou,zinnborel}. For an electric field
background, the weak-field expansion of the effective Lagrangian is a
non-alternating divergent series, which signals the presence of an
exponentially small imaginary part of the effective Lagrangian. 
Knowledge of the rate of divergence of the perturbation series can be
used to deduce the exponentially small imaginary part of the effective
Lagrangian, and therefore the pair production rate. 

This correspondence, between the divergent perturbative weak-field expansion and
the non-perturbative imaginary part, is well understood at the one-loop level. For
example, the weak-field expansions of the one-loop effective Lagrangians
(\ref{L1spinren}) and (\ref{L1scalren}), specialized to the constant electric field
case, are:
\begin{eqnarray}
{\cal L}^{(1)}_{\rm spin}(E)&=&-\frac{2m^4}{\pi^2}\left(\frac{eE}{m^2}\right)^4
\sum_{n=0}^\infty {(-1)^n 2^{2n} {\cal B}_{2n+4}\over (2n+4)(2n+3)(2n+2)}\left(
\frac{eE}{m^2}\right)^{2n}
\label{1lwspin}\\
{\cal L}^{(1)}_{\rm scal}(E)&=&\frac{m^4}{\pi^2}\left(\frac{eE}{m^2}\right)^4
\sum_{n=0}^\infty {(-1)^n 2^{2n}(2^{-2n-3}-1) {\cal B}_{2n+4}\over
(2n+4)(2n+3)(2n+2)}\left(
\frac{eE}{m^2}\right)^{2n}
\label{1lwscal}
\end{eqnarray}
where the coefficients involve the Bernoulli numbers ${\cal B}_{2n}$, which
alternate in sign and diverge factorially fast in magnitude
\cite{abramowitz,gradshteyn}:
\begin{eqnarray}
{\cal B}_{2n}=(-1)^{n+1}2 \frac{(2n)!}{(2\pi)^{2n}}\, \zeta(2n)
\label{bernoulli}
\end{eqnarray}
Here $\zeta(n)$ denotes the Riemann zeta function \cite{abramowitz}
\begin{eqnarray}
\zeta(n)=\sum_{k=1}^\infty \frac{1}{k^n}
\label{zeta}
\end{eqnarray}
which is exponentially close to 1 for large $n$. Also noting that \cite{abramowitz}
\begin{eqnarray}
(1-2^{1-n})\zeta(n)=\sum_{k=1}^\infty \frac{(-1)^{k+1}}{k^n}
\label{altzeta}
\end{eqnarray}
we see that the weak-field expansion coefficients in (\ref{1lwspin}) and
(\ref{1lwscal}) are
\begin{eqnarray}
a_{n}^{(1)\,{\rm spin}}& = &
\frac{2m^4}{\pi^2}\left(\frac{eE}{m^2}\right)^4\frac{\Gamma(2n+2)}{8}\,
\sum_{k=1}^\infty \frac{1}{(\pi k)^{2n+4}}
\label{1lsp}\\
a_{n}^{(1)\,{\rm scal}}& = &
\frac{m^4}{\pi^2}\left(\frac{eE}{m^2}\right)^4\frac{\Gamma(2n+2)}{8}\,
\sum_{k=1}^\infty \frac{(-1)^{k+1}}{(\pi k)^{2n+4}}
\label{1lsc}
\end{eqnarray}
There is a precise one-to-one correspondence (as will be reviewed in section 3)
between the terms in these large-order behaviors of the weak-field expansion
coefficients, and the terms in the non-perturbative expressions (\ref{fullimag}) and
(\ref{L1scalim}) for the imaginary parts of the one-loop effective Lagrangians. 

At the two-loop level, the situation is even more interesting. The first radiative
corrections to the Euler-Heisenberg Lagrangians (\ref{L1spinren}) and
(\ref{L1scalren}), describing the effect of an additional photon exchange in the
loop, were first studied in the seventies by Ritus \cite{ritus1,ritus2,ginzburg}.
Using the exact spinor and scalar propagators in a constant field found by Fock
\cite{fock} and Schwinger \cite{schwinger51}, and a proper-time cutoff as the UV
regulator, Ritus obtained the two-loop contribution
${\cal L}^{(2)}$ in terms of a certain two-parameter integral. 
This integral representation was then used in \cite{lebrit} for deriving, by an
analysis of the  analyticity properties of the integrand, a representation for the
imaginary part ${\rm Im}{\cal L}_{\rm spin}^{(2)}$, analogous to Schwinger's one-loop
formula (\ref{fullimag}).  Adding together the one-loop and the two-loop
contributions, the imaginary part reads, in the purely electric case for spinor QED,
\begin{eqnarray}
{\rm Im} {\cal L}_{\rm spin}^{(1)} (E) +
{\rm Im} {\cal L}_{\rm spin}^{(2)} (E) &=&  \frac{m^4}{8\pi^3}
\beta^2\,
\sum_{k=1}^\infty
\Bigl[
\frac{1}{k^2}
+\alpha\pi K_k(\beta)
\Bigr]
\,\exp\left[-\frac{\pi k}{\beta}\right]
\label{fullimag2loop}
\end{eqnarray}
where $\alpha=\frac{e^2}{4\pi}$ is the fine-structure constant. The coefficient
functions $K_k(\beta)$ appearing here were not obtained explicitly by \cite{lebrit}.
However, it was shown that they have small $\beta$ expansions of the following form:
\begin{eqnarray}
K_k(\beta) &=& -{c_k\over \sqrt{\beta}} + 1 + {\rm O}(\sqrt{\beta})
 \nonumber\\
c_1 = 0,\quad && \quad
c_k = {1\over 2\sqrt{k}}
\sum_{l=1}^{k-1} {1\over \sqrt{l(k-l)}},
\quad k \geq 2
\label{expK}
\end{eqnarray}
Note that, for $k\geq 2$, these expansions start with terms that
are singular in the limit of vanishing field $\beta\to 0$,
which seems to be at variance with the fact that these
coefficients have a direct physical meaning.
In \cite{lebrit} a physically intuitive solution
was offered to this dilemma. Its basic assumption is
that, if one would take into account all
contributions from even higher loop orders to the prefactor
of the $k$-th exponential, then one would find them
to exponentiate in the following way,
\begin{eqnarray}
\Bigl[
\frac{1}{k^2}
+\alpha\pi K_k\bigl({eE\over m^2}\bigr)
+\ldots
\Bigr]
\,\exp\left[-{k\pi m^2\over eE}\right]
=
\frac{1}{k^2}
\exp\left[-{k\pi m^2_{\ast}(k,E)\over eE}\right]
\label{shiftm}
\end{eqnarray}
Thus, it should be possible to absorb their effect
completely into
a field-dependent shift of the electron mass.
Using just the lowest order coefficients in the
small -- $\beta$ expansion of $K_k(\beta)$, those
given explicitly in (\ref{expK}), this
mass shift reads
\bear
m_{\ast}(k,E) &=&
m +\half\alpha kc_k\sqrt{eE}-\half\alpha keE/m
\label{massshift}
\ear
As shown in \cite{ritusmass,lebrit} these
contributions to the mass shift have a simple meaning in
the coherent tunneling picture:
The negative term can be interpreted as the total 
Coulomb energy of
attraction between opposite charges in
a coherent group; the positive one, which is present
only in the case $k\geq 2$,  represents the
energy of repulsion between like charges.

It is important to note that this interpretation of the mass
  shift requires the mass $m$ on the right hand side
 of (\ref{massshift}) to be the
  {\sl physical} renormalized electron mass of the vacuum theory.
  Only in this case the expansion of $K_k(\beta)$ has the form indicated
  in eqs.(\ref{expK}).
  It is an
  interesting side result of the Lebedev-Ritus
   analysis that the physical electron
  mass can be recognized from an inspection of the two-loop effective
  Lagrangian alone, without ever considering the one-loop electron
  mass operator.

Clearly, it would be of interest to understand in greater detail 
this prefactor series
$K_k(\beta)$. 
It is not even
clear from the work of Lebedev and Ritus whether or not this is a convergent
expansion. In an effort to learn more about this prefactor, the 
imaginary part of
the two-loop effective Lagrangian for a constant electric field background was
studied using Borel techniques \cite{ds1}. However, since no closed-form is known for
the two-loop weak-field expansion coefficients in the case of a background constant
electric field, this analysis was necessarily numerical. The leading,
and first sub-leading, growth rates of the expansion coefficients were deduced
numerically from the first 15 coefficients in the spinor QED case, using a
brute-force expansion of the integral representation (a double integral) of the
effective Lagrangian. This was sufficient to indicate that there are power-law
corrections to the leading factorial growth rate, in contrast to the one-loop
results (\ref{1lsp}) and (\ref{1lsc}) which only have {\it exponential} corrections
to the leading growth rate. Using Borel dispersion relations, this indicates the
presence of a prefactor series, as in the Ledebev-Ritus
result (\ref{fullimag2loop}). In this way, a numerical estimate was obtained for the
next term in the expansion of the prefactor $K_1(\beta)$ of the leading exponential
term ($k=1$). This also provided an independent confirmation (albeit numerical) that
the expansion proceeds in powers of $\sqrt{\beta}$. However, it
was prohibitively difficult to go to much higher order, and so it was only possible
to deduce one more term in the prefactor expansion. 

However, the case of a constant electric field background is not the 
simplest case
one can study in this context. As shown in part I \cite{sd1}, for a
constant self-dual Euclidean background, the two-loop effective Lagrangians, for
both spinor and scalar QED, simplify dramatically, to such a degree that there are
simple closed-form expressions in terms of the digamma function \cite{sdletter}.  
Such a constant  background field satisfies 
\bear
F_{\mu\nu}=\tilde
F_{\mu\nu}\equiv \half\varepsilon_{\mu\nu\alpha\beta}F^{\alpha\beta}.
\label{defsd}
\ear
This has the consequence that
\begin{eqnarray}
F^2 = - f^2\Eins
\label{f}
\end{eqnarray}
In Minkowski space, the self-duality condition
(\ref{defsd}) requires either $\bf E$ or $\bf B$ to
be complex. This does not imply that such backgrounds are 
devoid of physical meaning; 
rather, the effective action in such a background
should be understood in terms of helicity projections 
\cite{duff,sd1}, and contains information on the photon
amplitudes with all equal helicities (see part I).

As shown in part I, for such a self-dual background, the two-loop
spinor QED and scalar QED renormalized effective Lagrangians are:
\bear
{\cal L}_{\rm spin}^{(2)(SD)}(\kappa)
&=&
-2\alpha \,{m^4\over (4\pi)^3}\frac{1}{\kappa^2}\left[
3\xi^2 (\kappa)
-\xi'(\kappa)\right]
\label{2lspintro}\\
{\cal L}_{\rm scal}^{(2)(SD)}(\kappa)
&=&
\alpha \,{m^4\over (4\pi)^3}\frac{1}{\kappa^2}\left[
{3\over 2}\xi^2 (\kappa)
-\xi'(\kappa)\right]
\label{2lscintro}
\ear
Here we have defined the convenient dimensionless parameter
\begin{eqnarray}
\kappa\equiv \frac{m^2}{2e\sqrt{f^2}}
\label{kappa}
\end{eqnarray}
as well as the important function
\bear
\xi(x)\equiv &=& -x\Bigl(\psi(x)-\ln(x)+{1\over 2x}\Bigr)
\label{defxi}
\ear
with $\psi$ being the digamma function
$\psi(x)=\Gamma^\prime(x)/\Gamma(x)$ \cite{abramowitz,gradshteyn}.

In \cite{sd1}, the parameter $\kappa$, defined in (\ref{kappa}), was taken to be
real. In Minkowski space, this corresponds to taking the magnetic field $B$ to
be real, and the electric field $E$ to be imaginary -- so we refer to this case as
the self-dual 'magnetic' case. In this case, the two-loop effective Lagrangians
(\ref{2lspintro}) and (\ref{2lscintro}) are real. However, we could also
analytically continue $\kappa\to i\kappa$, which in the Minkowski language
corresponds to taking $E$ to be real and $B$ to be imaginary. Thus, we refer to this
case as the self-dual 'electric' case. For this background, with $\kappa$ imaginary,
the two-loop effective Lagrangians (\ref{2lspintro}) and (\ref{2lscintro}) acquire
exponentially small imaginary parts. As we shall see, these imaginary parts share
many of the properties of the imaginary parts in (\ref{fullimag}) and
(\ref{L1scalim}), for the case of a constant electric field background. 

The obvious advantage of the self-dual case is that since there are simple
closed-form expressions for the two-loop effective Lagrangians (\ref{2lspintro}) and
(\ref{2lscintro}), it is possible to study the imaginary part in much greater
detail. In this paper, we analyze these imaginary parts, and compare the direct
approach with the Borel dispersion relation approach. We are able to find the entire
prefactor series for the imaginary parts. Since the spinor QED and scalar
QED cases are so similar for a self-dual background, to avoid repetition, we
concentrate on just one of them -- the scalar QED case. In section 2 we present the
details of the weak- and strong-field expansions of the one- and two-loop effective
Lagrangians for scalar QED in a constant self-dual background. In section 3 we apply
Borel summation techniques to the weak-field expansion, and use Borel dispersion
relations to compute the imaginary part of the effective Lagrangian when the field
strength is analytically continued $\kappa\to i \kappa$. Section 4 contains some
concluding remarks, and an Appendix contains the details of the derivation of the
large-order behavior of the two-loop weak-field expansion coefficients.

\section{Weak-field and strong-field limits (scalar QED)}
\label{weakstrong}
\renewcommand{\theequation}{2.\arabic{equation}}
\setcounter{equation}{0}

In this section we discuss the weak- and strong-field limits of the
on-shell renormalized one- and two-loop effective actions for scalar QED in a
constant Euclidean self-dual background.  We begin by
recalling (\ref{kappa}) that $\kappa=\frac{m^2}{2e\sqrt{f^2}}$, where
$f$ is defined in (\ref{f}), so that "weak-field" means large $\kappa$, and
"strong-field" means small $\kappa$. We first review the one-loop case, and then turn
to the two-loop case.

\subsection{One Loop}
In the self-dual (`SD') case, the integral representation (\ref{L1scalren})
for the renormalized one-loop effective Lagrangian becomes

\begin{eqnarray}
{\cal L}_{\rm
scal}^{(1)(SD)}(\kappa)=\frac{m^4}{(4\pi)^2}\frac{1}{4\kappa^2}\int_0^\infty
\frac{dt}{t^3}\, e^{-2\kappa t}\left[\frac{t^2}{\sinh^2(t)}-1+\frac{t^2}{3}\right]
\label{1lsclag}
\end{eqnarray}
The weak-field (large $\kappa$) expansion of this proper-time integral
representation can be derived using the Taylor expansion \cite{abramowitz,gradshteyn}
\bear
\Bigl({x\over\sinh(x)}\Bigr)^2
&=&
-\sum_{k=0}^{\infty}
{(2k-1)2^{2k}\over (2k)!}{\cal B}_{2k}x^{2k}
\label{taylorz/sinh}
\ear
This leads to the weak-field expansion
\begin{eqnarray}
{\cal L}_{\rm scal}^{(1)(SD)}(\kappa)
=\frac{m^4}{(4\pi)^2}\,\sum_{n=2}^\infty c_n^{(1)}\frac{1}{\kappa^{2n}}
\label{1lscweak}
\end{eqnarray}
where the expansion coefficients are (for $n\geq 2$):
\begin{eqnarray}
c_n^{(1)}=- \frac{{\cal B}_{2n}}{2n(2n-2)}
\label{cn1}
\end{eqnarray}
The leading term in the weak-field expansion (\ref{1lscweak}) is
\begin{eqnarray}
{\cal L}_{\rm scal}^{(1)(SD)}(\kappa)
\sim \frac{m^4}{(4\pi)^2}\,\frac{1}{240\kappa^{4}}
\label{1lscweakleading}
\end{eqnarray}

To derive the strong-field (small $\kappa$) expansion, it is convenient to express
(\ref{1lsclag}) in terms of special functions as
\begin{eqnarray}
{\cal L}_{\rm scal}^{(1)(SD)}(\kappa)= \frac{m^4}{(4\pi)^2}\frac{1}{\kappa^2}
\left[-\frac{1}{12}\ln \kappa +\zeta^\prime(-1)+\Xi(\kappa)\right]
\label{1lscll}
\end{eqnarray}
where the function $\Xi(\kappa)$ is defined as \cite{barnes}
\begin{eqnarray}
\Xi(\kappa)\equiv -\kappa\,\ln \Gamma(\kappa) +\frac{\kappa^2}{2} \ln \kappa
-\frac{\kappa^2}{4}-\frac{\kappa}{2} +\int_0^\kappa dy \, \ln\Gamma(y)
\label{Xidef}
\end{eqnarray}
and $\zeta^\prime(-1)\approx -0.16542$. Note that this function $\Xi(\kappa)$ is
simply related to the function $\xi(\kappa)$, defined in (\ref{defxi}), which
appears in the two-loop expressions (\ref{2lspintro}) and (\ref{2lscintro}) by
$\Xi^\prime(\kappa)=\xi(\kappa)$.

\begin{figure}[ht]
\centerline{\includegraphics[scale=1]{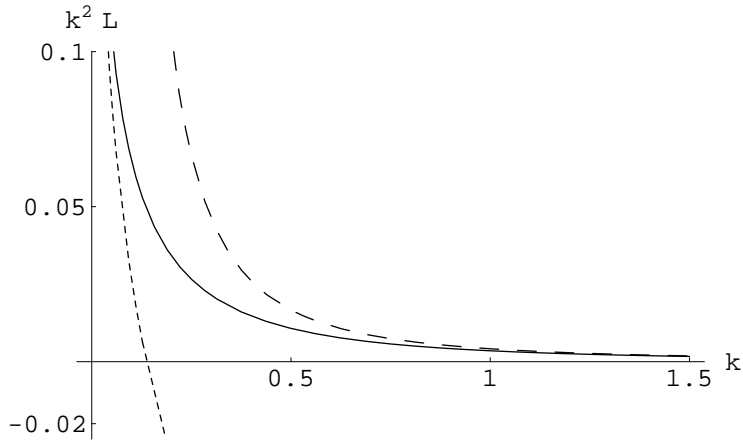}}
\caption{This plot shows the exact one-loop scalar QED effective
Lagrangian (\protect{\ref{1lsclag}}) [solid curve] in comparison with the
leading weak-field (large $\kappa$) [long-dashed curve] and strong-field
(small $\kappa$) [short-dashed curve] contributions
(\protect{\ref{1lscweakleading}}) and
(\protect{\ref{1lscstrongleading}}), respectively. We have multiplied
through each function by a common factor of
$(4\pi)^2\kappa^2/m^4$ for ease of presentation.}
\label{f1}
\end{figure}

Thus, using the Taylor expansion of $\ln \Gamma(x)$,
\begin{eqnarray}
\ln \Gamma(x)=-\ln x-\gamma x +\sum_{n=2}^\infty \frac{(-1)^n
\zeta(n)}{n} x^n
\label{loggamma}
\end{eqnarray}
the strong-field (small $\kappa$) expansion of the one-loop effective action
(\ref{1lsclag}) is
\begin{eqnarray}
{\cal L}_{\rm
scal}^{(1)(SD)}(\kappa)=\frac{m^4}{(4\pi)^2\kappa^2}
\left[\left(-\frac{1}{12}+\frac{\kappa^2}{2}\right)\ln
\kappa +\zeta^\prime(-1)+\frac{\kappa}{2}+
(\frac{\gamma}{2}-\frac{1}{4})\kappa^2-
\sum_{n=2}^\infty \frac{(-1)^n \zeta(n)}{(n+1)}\,
\kappa^{n+1}\right]
\non\\
\label{1lscstrong}
\end{eqnarray}
The leading strong-field behavior in (\ref{1lscstrong}) is
\begin{eqnarray}
{\cal L}_{\rm scal}^{(1)(SD)}(\kappa) \sim \,
\frac{m^4}{(4\pi)^2}\,\frac{1}{\kappa^2}\left[
-\,\frac{1}{12}\ln\kappa+\zeta^\prime(-1)\right]
\label{1lscstrongleading}
\end{eqnarray}
Figure \ref{f1} shows a plot of ($\kappa^2$ times) the exact one-loop
effective action (\ref{1lsc}), compared with ($\kappa^2$ times) the
leading weak- and strong-field behaviors. (We multiply by the common
factor of
$\kappa^2$ for clarity of presentation over a reasonable range of
$\kappa$). From Figure \ref{f1}, it is clear that the leading behaviors
(\ref{1lscweakleading}) and (\ref{1lscstrongleading}) correctly capture
the extremes, but fail to connect well in the intermediate region.
(See \cite{chiral} for a simple approximate interpolation method to
connect the weak- and strong-field extremes).

\begin{table}
\begin{centering}
\begin{tabular}{|c|c|c|c|c|}
\hline
n&$c_n^{(1)}$&$N[|c_n^{(1)}|]$&$c_n^{(2)}$&$N[|c_n^{(2)}|]$\cr\hline 
2&$\frac{1}{240}$&0.00416667&$\frac{3}{128\,{\pi
}^2}$&0.00237472\cr\hline
3&$-\frac{1}{1008}$&0.000992063&$-\frac{13}{1920\,{\pi}^2}$
&0.000686029\cr\hline 
4&$\frac{1}{1440}$&0.000694444&$\frac{67}{12800\,{\pi }^2}$&
$0.000530353$\cr\hline 
5&$-\frac{1}{1056}$&0.00094697&$\frac{-611}{80640\,{\pi
}^2}$&$0.000767699$\cr\hline
6&$\frac{691}{327600}$&0.00210928&$\frac{3269351}{186278400\,{\pi}^2}$
&$0.00177828$\cr\hline 
7&$-\frac{1}{144}$&0.00694444&$\frac{-684779}{11531520\,{\pi}^2}$
&$0.00601678$\cr\hline 
8&$\frac{3617}{114240}$&0.0316614&$\frac{42467137}{153753600\,{\pi}^2}$
&$0.0279852$\cr\hline 
9&$-\frac{43867}{229824}$&0.190872&$\frac{-2783806241}{1646701056\,{\pi
}^2}$ &$0.171287$\cr\hline  
10&$
\frac{174611}{118800}$&1.46979&$\frac{308416614839}{23412280320\,{\pi
}^2}$ &$1.33473$\cr\hline  
11&$-\frac{77683}{5520}$&14.073&$\frac{-14229172307981}{111740428800\,{\pi
}^2}$&$12.9024$\cr\hline  
12&$\frac{236364091}{1441440}$&163.978&
$\frac{20984465589542501429}{14032363048704000\,{\pi}^2}$&
$151.519$\cr\hline 
13&$-\frac{657931}{288}$&2284.48&$\frac{-7699261058623757}{367147123200\,{\pi
}^2}$&2124.76\cr\hline 
14&$\frac{3392780147}{90480}$&37497.6&
$\frac{8895454541748900227}{25700298624000\,{\pi}^2}$&35069.6\cr\hline 
15&$-\frac{1723168255201}{2406096}$&716168&
$\frac{-114354778628165307771811}{17216630048217600\,{\pi}^2}$
&672987\cr\hline
16&$\frac{7709321041217}{489600}   $&$ 1.57462 \times 10^7 
$&$\frac{460780386050723601316078067} {3142221310964736000\,{\pi }^2} 
$&$1.48579 \times {10}^7 $\cr\hline
\end{tabular}
\caption{The first 15 one- and two-loop weak-field expansion
coefficients, $c_n^{(1)}$ and $c_n^{(2)}$ appearing in
(\protect{\ref{cn1}}) and (\protect{\ref{cn2}}), together with their
numerical magnitudes. Note that in each case the expansion coefficients
alternate in sign and grow factorially fast in magnitude. Furthermore,
the leading growth rate at one- and two-loop is the same.}
\end{centering}
\label{t1}
\end{table}

From (\ref{cn1}) and (\ref{bernoulli}), it is clear that the weak-field
expansion (\ref{1lscweak}) is a divergent expansion. The first 15
expansion coefficients
$c_n^{(1)}$ are tabulated in Table \ref{t1}, together with their magnitudes. Notice
that the coefficients alternate in sign (here $\kappa$ is real), and their magnitude
eventually grows very fast. In fact, the leading growth rate of the coefficients at
large $n$ is factorial:
\begin{eqnarray}
c_n^{(1)}\sim  2
\frac{(-1)^{n}}{(2\pi)^{2n}}\,\Gamma(2n-1)\left[1+
O\left(\frac{1}{n}\right)\right]
\label{1lscgrowth}
\end{eqnarray}

\begin{figure}[ht]
\centerline{\includegraphics[scale=1]{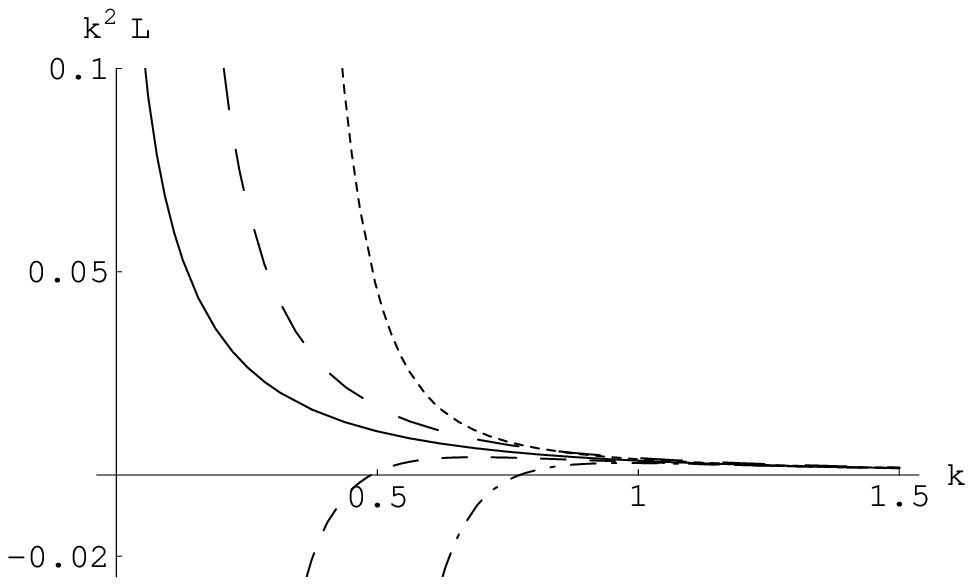}}
\caption{This plot shows the exact one-loop scalar QED effective
Lagrangian (\protect{\ref{1lsclag}}) [solid curve] in comparison with
successive partial sums of the weak-field (large $\kappa$) expansion
(\protect{\ref{1lscweak}}). The partial sum is shown for one term
[long-dashed-curve], two terms [medium-dashed curve], three terms
[short-dashed curve], and four terms [dot-dash curve]. We have
multiplied through each function by a common factor of
$(4\pi)^2\kappa^2/m^4$ for ease of presentation.}
\label{f2}
\end{figure}

Figure \ref{f2} shows a plot comparing the exact one-loop result
(\ref{1lsclag}) with successive truncations of the weak-field series
expansion in (\ref{1lscweak}). (Note once again that these plots have been
multiplied by a common factor of $\kappa^2$). We see that the successive truncations
form an envelope around the exact result. This is typical behavior for an
alternating asymptotic expansion.

To probe more deeply the asymptotic nature of the weak-field expansion we
consider the optimal asymptotic approximation \cite{carlbook,west} given by
truncating the series in such a way that the error is minimized. (More
sophisticated estimation techniques can be applied to the series, but the simplest
one is sufficient here). To find the optimal truncation point of an asymptotic series
\begin{eqnarray}
f(x)=\sum_n (-1)^n c_n x^n
\label{eg}
\end{eqnarray}
we evaluate the terms $c_n x^n$ and find the minimum value, say
$c_{N_0}x^{N_0}$. Then the optimal asymptotic approximation consists of
keeping terms up to $n=N_0-1$. In this way the $N_0^{th}$ term, which is
a measure of the truncation error, is minimized. Clearly, the truncation
point depends on the value of expansion parameter $x$.

\begin{table}
\begin{centering}
\begin{tabular}{|c|c|c|c|c|}
\hline
n&$\kappa=0.5$&$\kappa=1$&$\kappa=1.5$&$\kappa=2$\\ \hline 
2&0.0666667&0.00416667&0.000823045&0.000260417\\ \hline
3&0.0634921&0.000992063&0.0000870947&0.000015501\\ \hline
4&0.177778&0.000694444&0.0000270961& $2.71267 \times {10}^{-6}$\\ \hline
5&0.969697&0.00094697&0.0000164219&$9.24775 \times {10}^{-7}$\\ \hline
6&8.63961&0.00210928&0.0000162569&$5.14961 \times {10}^{-7}$\\ \hline
7&113.778&0.00694444&0.0000237881&$4.23855 \times {10}^{-7}$\\ \hline
8&2074.96&0.0316614&0.0000482026&$4.83115 \times {10}^{-7}$\\ \hline
9&50036&0.190872&0.000129152&$7.28119 \times {10}^{-7}$\\ \hline
10&$1.54119 \times {10}^6$&1.46979&0.000442008&$1.4017 \times
{10}^{-6}$\\ \hline 
11&$5.90265 \times {10}^7$&14.073&0.00188096&$3.35527 \times
{10}^{-6}$\\ \hline \hline
$n_{\rm min}$ &3&4&6&7\\ \hline
$N_0=\frac{1}{2}+\pi\kappa\,\exp(\frac{1}{4\pi\kappa})$& 2.34179& 3.90181
& 5.46914& 7.03823\\ \hline\hline
optimal partial sum & 0.0666667  & 0.0031746  & 0.000746625  &
0.000247219 \\ \hline  
exact value & 0.0429404 & 0.00351739 & 0.000753997
& 0.000247005 \\ \hline\hline
$\%$ error& 55.2539&9.74548&0.977739&0.0865725\\ \hline
\end{tabular}
\label{t2}
\caption{This table shows the magnitudes of the terms
$c_n^{(1)}/\kappa^{2n}$ in the one-loop weak-field expansion
(\protect{\ref{1lscweak}}), for four different values of $\kappa$. In
each column, note that the terms decrease in magnitude and then
increase in magnitude. The order of the minimum term is
$n_{\rm min}$, and this is compared with the estimate $N_0$ of the
order of the minimum term given in (\protect{\ref{trunc}}). The optimal
partial sum is the partial sum up to the $(n_{\rm min}-1)^{th}$ term.
This is compared with the exact value and the fractional error is shown
as a percentage.}
\end{centering}
\end{table}

For the specific case of the weak field expansion (\ref{1lscweak}), we
define the remainder as
\begin{eqnarray}
R_N(\kappa)&=&-\frac{m^4}{(4\pi)^2}\,\frac{{\cal
B}_{2N}}{2N(2N-2)}\,\frac{1}{\kappa^{2N}}\nonumber\\
&\sim & -\frac{m^4}{(4\pi)^2}\,2
\Gamma(2N-1)\,e^{-N\ln(2\pi\kappa)^2}
\label{remainder}
\end{eqnarray}
To find the optimal truncation point $N_0$, we minimize $|R_N(\kappa)|$
with respect to $N$. The magnitude of this remainder is minimized when
$\ln(2\pi\kappa)=\psi(2N-1)\approx \ln(2N-1)-1/(2(2N-1))+...$. Thus, a
simple estimate of the point at which the remainder is minimized is given
by 
\begin{eqnarray}
N_0\approx \frac{1}{2}+\pi\kappa \, \exp\left(\frac{1}{4\pi\kappa}\right)
\label{trunc}
\end{eqnarray}
In Table 2 we compare this optimal asymptotic approximation with the
exact answer, for various values of $\kappa$. Note that for $\kappa
< 1$ (i.e., "strong field", where $f>\frac{m^2}{2e}$), the optimal truncation
involves keeping just the first term in the weak-field expansion. Keeping more terms
in the series makes the series estimate worse.

\subsection{Two Loop}

At two-loop, for scalar QED in a constant Euclidean selfdual background,
the on-shell renormalized effective Lagrangian is given by (\ref{2lscintro}).
The weak-field (large $\kappa$) expansion of this result is
\bear
{\cal L}^{(2)(SD)}_{\rm scal}(\kappa)= \alpha\pi\,
{m^4\over (4\pi)^2}\, 
\sum_{n=2}^{\infty} c^{(2)}_n \frac{1}{\kappa^{2n}}
\label{2lscweak}
\end{eqnarray}
where the two-loop expansion coefficients are (for $n\geq 2$):
\begin{eqnarray}
c^{(2)}_n =
{1\over (2\pi)^2}\biggl\lbrace
\frac{2n-3}{2n-2}\,{\cal B}_{2n-2}
+\frac{3}{2}\sum_{k=1}^{n-1}
{{\cal B}_{2k}\over 2k}
{{\cal B}_{2n-2k}\over (2n-2k)}
\biggr\rbrace
\label{cn2}
\ear
\begin{figure}[ht]
\centerline{\includegraphics[scale=1]{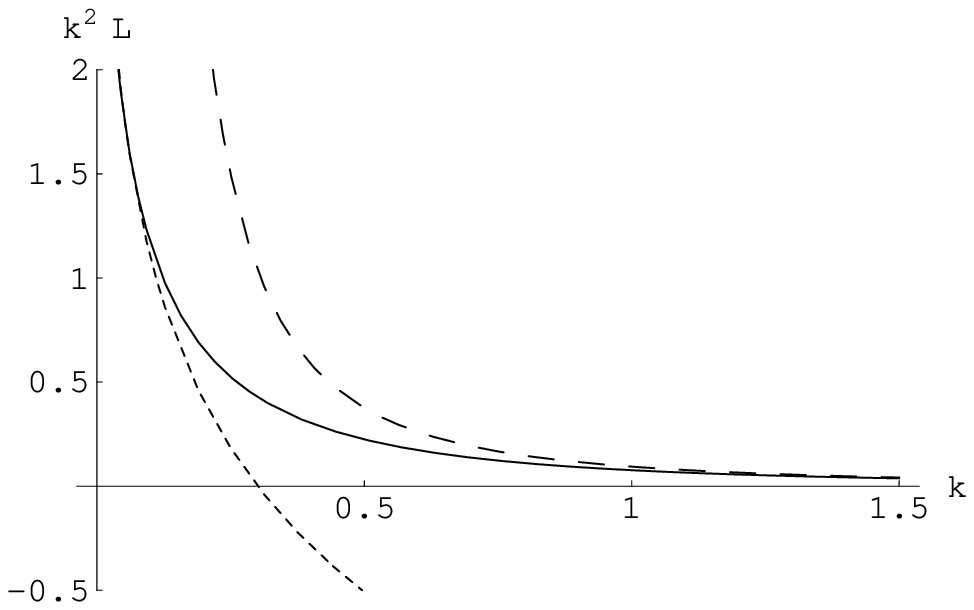}}
\caption{This plot shows the exact two-loop scalar QED effective
Lagrangian (\protect{\ref{2lscintro}}) [solid curve] in comparison with the
leading weak-field (large $\kappa$) [long-dashed curve] and strong-field
(small $\kappa$) [short-dashed curve] contributions
(\protect{\ref{2lscweakleading}}) and
(\protect{\ref{2lscstrongleading}}), respectively. We have multiplied
through each function by a common factor of
$(4\pi)^2\kappa^2/(\alpha \pi m^4)$ for ease of presentation.}
\label{f3}
\end{figure}

This weak-field expansion follows directly from (\ref{2lscintro}) and
(\ref{defxi}), together with the asymptotic (large $x$) expansion of the digamma
function
\cite{abramowitz,gradshteyn}:
\begin{eqnarray}
\psi(x)\sim \ln x-\frac{1}{2x}-\sum_{k=1}^\infty \frac{{\cal B}_{2k}}{2k
\,x^{2k}} 
\label{psiasymptotic}
\end{eqnarray}
The leading term in the weak-field expansion (\ref{2lscweak}) is 
\bear
{\cal L}^{(2)(SD)}_{\rm scal}(\kappa)\sim \alpha\pi\,
{m^4\over (4\pi)^2}\,  \frac{3}{128\,{\pi
}^2\kappa^{4}}
\label{2lscweakleading}
\end{eqnarray}

The strong-field (small $\kappa$) expansion follows directly from 
(\ref{2lscintro}) and (\ref{defxi}), together with the small argument expansion
of the digamma function \cite{abramowitz,gradshteyn}:
\begin{eqnarray}
\psi(x)\sim -\frac{1}{x}-\gamma +\sum_{k=2}^\infty (-1)^k \zeta(k) x^{k-1}
\label{psitaylor}
\end{eqnarray}
This leads to:
\begin{eqnarray}
{\cal L}_{\rm scal}^{(2)(SD)}(\kappa)&=& \frac{m^4
\alpha}{(4\pi)^3}\,\frac{1}{\kappa^2}\left[ \,\left(\gamma+\ln
\kappa\right)\left\{-1+\frac{3}{2}\kappa+
\frac{3}{2}\kappa^2(\gamma+\ln\kappa)-3\sum_{n=2}^\infty (-1)^n \zeta(n)
\kappa^{n+1}\right\}-\frac{5}{8}\right.\nonumber\\
&&\hskip -2cm \left.+\frac{\pi^2}{3}\kappa -\sum_{n=2}^\infty (-1)^n
\left(\frac{3}{2}\zeta(n)+(n+1)\zeta(n+1)\right)\kappa^n 
+\frac{3}{2}\sum_{n=4}^\infty\sum_{l=2}^{n-2}(-1)^n \zeta(n-l)\zeta(l)
\kappa^n\right] 
\label{2lscstrong}
\end{eqnarray}
The leading strong-field behavior is
\begin{eqnarray}
{\cal L}_{\rm scal}^{(2)(SD)} \sim \frac{m^4
\alpha}{(4\pi)^3}\,\frac{1}{\kappa^2}\left[-\ln\kappa
-\frac{5}{8}-\gamma\right]
\label{2lscstrongleading}
\end{eqnarray}
Figure \ref{f3} shows a plot of ($\kappa^2$ times) the exact expression
(\ref{2lscintro}), in comparison with ($\kappa^2$ times) these leading
weak-field and strong-field behaviors, (\ref{2lscweakleading}) and
(\ref{2lscstrongleading}), respectively. This is completely analogous 
to the one-loop case (see Fig.\ref{f1}). The extreme behavior for large and small
$\kappa$ is well described, but they do not connect well in the intermediate region.

\begin{figure}[ht]
\centerline{\includegraphics[scale=1]{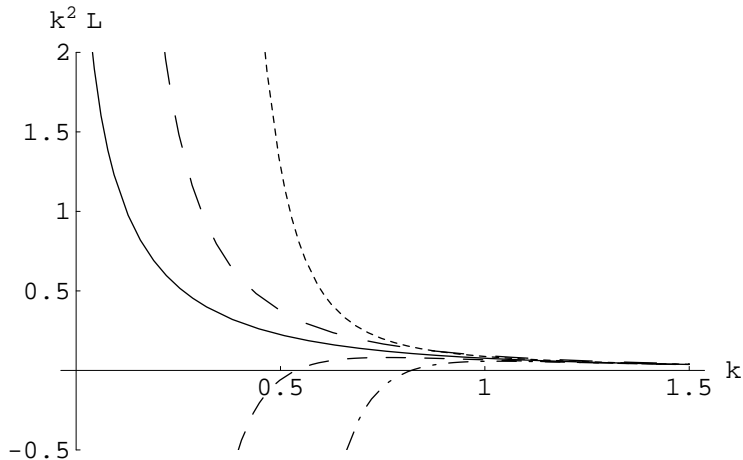}}
\caption{This plot shows the exact two-loop scalar QED effective
Lagrangian (\protect{\ref{2lscintro}}) [solid curve] in comparison with
successive partial sums of the weak-field (large $\kappa$) expansion
(\protect{\ref{2lscweak}}). The partial sum is shown for one term
[long-dashed-curve], two terms [medium-dashed curve], three terms
[short-dashed curve], and four terms [dot-dash curve]. We have multiplied
through each function by a common factor of
$(4\pi)^2\kappa^2/(\alpha \pi m^4)$ for ease of presentation.}
\label{f4}
\end{figure}

The two-loop weak-field expansion (\ref{2lscweak}) is a divergent asymptotic
expansion; it is simply based on the asymptotic expansion
(\ref{psiasymptotic}) of the digamma function $\psi(\kappa)$. Figure
\ref{f4} shows a comparison of the exact two-loop expression (\ref{2lscintro})
with successive truncations of the weak-field expansion in
(\ref{2lscweak}) (Once again, a common extra factor of $\kappa^2$ is
included for clarity of plotting). The pattern is much the same as in the
one-loop case shown in Fig. \ref{f2}, and is characteristic of an asymptotic
expansion.

The two-loop expansion coefficients (\ref{cn2}) alternate in sign and grow
factorially fast in magnitude, as shown in Table 1. In fact, the precise leading
growth rate is 
\begin{eqnarray}
c_n^{(2)}\sim  2
\frac{(-1)^{n}}{(2\pi)^{2n}}\,\Gamma(2n-1)\left[1+
O\left(\frac{1}{n}\right)\right]
\label{2lscgrowth}
\end{eqnarray}
which is exactly the same as the growth rate of the one-loop cofficients
$c_n^{(1)}$ in (\ref{1lscgrowth}). This equality between the one-loop and
two-loop leading growth rates can even be seen in the first 15 coefficients shown in
Table \ref{t1}. This equality will become significant in the following when we
discuss the Borel summation properties of the weak-field expansion. It also means
that the estimate of the optimal truncation point is exactly the same as in the
one-loop case, so we do not repeat the argument.

\section{Borel analysis}
\label{borelanal}
\renewcommand{\theequation}{3.\arabic{equation}}
\setcounter{equation}{0}

The weak-field expansions (\ref{1lscweak}) and (\ref{2lscweak}) at one-
and two-loop, respectively, are divergent asymptotic expansions. In this
section we apply the technique of Borel summation to obtain approximate
expressions for the corresponding one- and two-loop effective Lagrangians.
This enables us to test the Borel method with great accuracy because we
are in the unusual position of having the explicit closed-form 
expressions (\ref{1lsclag}) and (\ref{2lscintro}) for the
effective Lagrangians themselves, as well as the explicit expressions
(\ref{cn1}) and (\ref{cn2}) for the expansion coefficients to all
orders. We also consider the issue of analytic continuation in
the field strength, and compare the Borel results with the exact
results. 

Strictly speaking, it is, of course, unnecessary to perform a
Borel resummation in these cases, since we know the exact closed-form
answers. However, such knowledge of the exact answer is rare, and
our aim in this section is to probe the power, as well as the
limitations, of the Borel approach in a case where a precise
comparison can be made.

Note of course that divergent behavior is not a bad thing; it is
in fact a generic behavior in perturbation theory, as is illustrated in
numerous and diverse examples in both quantum field theory and quantum
mechanics \cite{dyson,arkady,leguillou,zinnborel}. It is well known that knowledge
of the divergence rate of high orders of perturbation theory can be used to extract
information about non-perturbative decay and tunneling rates, thereby
providing a bridge between perturbative and non-perturbative physics. 

To begin, we review very briefly some basics of Borel summation
\cite{hardy,carlbook,zinnborel,thooftborel}. Consider an asymptotic series
expansion of some function
$f(g)$
\begin{eqnarray}
f(g)\sim \sum_{n=0}^\infty \, a_n\, g^n
\label{exp}
\end{eqnarray}
where $g\to 0^+$ is a small dimensionless perturbation expansion parameter.
In an extremely broad range of physics applications \cite{leguillou} one finds
that perturbation theory leads not to a convergent series but to a divergent
series in which the expansion coefficients $a_n$ have leading large-order
behaviour
\begin{eqnarray}
a_n\sim (-1)^n \rho^n \Gamma(\mu\, n+\nu)  \qquad\qquad (n\to\infty)
\label{general}
\end{eqnarray}
for some real constants $\rho$, $\mu>0$, and $\nu$. When $\rho>0$, the
perturbative expansion coefficients $a_n$ alternate in sign and their
magnitude grows factorially, just as in the weak-field expansions studied
here: see (\ref{1lscgrowth}) and (\ref{2lscgrowth}). Borel summation is
a useful approach to this case of a divergent, but alternating series.
Non-alternating series must be treated somewhat differently (see below).

To motivate the Borel approach, consider the classic example :
$a_n=(-1)^n \rho^n n!$, and $\rho>0$. The series (\ref{exp}) is clearly
divergent for any value of the expansion parameter $g$. Write
\begin{eqnarray}
f(g)&\sim & \sum_{n=0}^\infty (-1)^n (\rho g)^n \, \int_0^\infty dt\, t^n\,
e^{-t}\nonumber\\
&\sim&
\frac{1}{\rho g}\, \int_0^\infty \,dt\, \left({1\over 1+t}\right) \,
\exp\left[- \frac{t}{\rho g}\right]
\label{borel}
\end{eqnarray}
where we have formally interchanged the order of summation and integration.
The final integral, which is convergent for all $g >0$, is {\it defined} to
be the sum of the divergent series. To be more precise \cite{hardy,carlbook}, the
formula (\ref{borel}) should be read backwards: for $g\to 0^+$, we can use
Laplace's method to make an asymptotic expansion of the integral, and we
obtain the asymptotic series in (\ref{exp}) with expansion coefficients
$a_n=(-1)^n \rho^n n!$. This example captures the essence of the Borel
method.

For a non-alternating series, such as $a_n=\rho^n n!$, we need $f(-g)$. The
particular Borel integral (\ref{borel}) is \cite{hardy,carlbook} an analytic
function of $g$ in the cut $g$ plane: $|{\rm arg}(g)|<\pi$. So a
dispersion relation (using the discontinuity across the cut along the
negative $g$ axis) can be used to {\it define} the imaginary part of
$f(g)$ for negative values of the expansion parameter:
\begin{eqnarray}
{\rm Im} \left[f(-g)\right]\sim\frac{\pi}{\rho g}\exp[-\frac{1}{\rho g}]
\label{imag}
\end{eqnarray}
The imaginary contribution (\ref{imag}) is non-perturbative (it clearly does
not have an expansion in positive powers of $g$) and has important physical
consequences. Note that (\ref{imag}) is consistent with a principal parts
prescription for the pole that appears on the $s>0$ axis if we make the formal
manipulations as in (\ref{borel}):
\begin{eqnarray}
\sum_{n=0}^\infty \rho^n n! \, g^n \sim \frac{1}{\rho g}\int_0^\infty
dt\,\left(\frac{1}{1-t}\right)\exp\left[-\frac{t}{\rho g}\right]
\label{non}
\end{eqnarray}

Similar formal arguments can be applied to the case when the expansion
coefficients have leading behaviour (\ref{general}). Then the leading Borel
approximation is
\begin{eqnarray}
f(g)\sim \frac{1}{\mu}\, \int_0^\infty \frac{dt}{t} \,
\left(\frac{1}{1+t}\right)
\left(\frac{t}{\rho g}\right)^{\nu/\mu}\,
\exp\left[-\left(\frac{t}{\rho g}\right)^{1/\mu}\right]
\label{genborel}
\end{eqnarray}
For the corresponding non-alternating case, when $g$ is negative, the leading
imaginary contribution is
\begin{eqnarray}
{\rm Im} \left[f(-g)\right]\sim\frac{\pi}{\mu}\left(\frac{1}{\rho g}
\right)^{\nu/\mu}
\exp\left[-\left(\frac{1}{\rho g}\right)^{1/\mu}\right]
\label{genimag}
\end{eqnarray}
Note the separate meanings of the parameters $\rho$, $\mu$ and $\nu$ that
appear in the formula (\ref{general}) for the leading large-order growth of
the expansion coefficients. The constant $\rho$ clearly combines with $g$ as
an effective expansion parameter. The power of the exponent in
(\ref{genborel}) and (\ref{genimag}) is determined by $\mu$, while the
power of the prefactor in (\ref{genborel}) (\ref{genimag}) is determined
by the ratio $\frac{\nu}{\mu}$.

To illustrate this correspondence, between the perturbative expansion coefficients
(\ref{general}) and the imaginary part (\ref{genimag}), it is now a straightforward
exercise to check that for the one-loop electric field background, the weak-field
expansion coefficients in (\ref{1lsp}) and (\ref{1lsc}) correspond to the imaginary
part of the respective effective Lagrangians given in (\ref{fullimag}) and
(\ref{L1scalim}).

These formulas (\ref{genborel}) and (\ref{genimag})
are somewhat formal, as they are based on assumed analyticity properties of
the function $f(g)$. These can be rigorously proved when the expansion
coefficients are precisely given by the gamma function form in
(\ref{general}). But things are less clear when this expression gives
just the {\it approximate} large order behavior of the coefficients. The
Borel dispersion relations could be complicated by the appearance of
additional poles and/or cuts in the complex $g$ plane, signalling new
physics, such as, for example, renormalons \cite{thooftborel,beneke}. Also, the
expression (\ref{genimag}) assumes a principal parts prescription for the poles. Our
analysis below will show that these assumptions are justified for the one- and
two-loop effective Lagrangians (\ref{1lsclag}) and (\ref{2lscintro}) for scalar QED
in a constant Euclidean self-dual background. The same is true for the spinor case,
but we do not repeat the analysis here: it is very similar.

\subsection{Borel Analysis of One Loop Case (Scalar QED)}

\begin{figure}[ht]
\centerline{\includegraphics[scale=1]{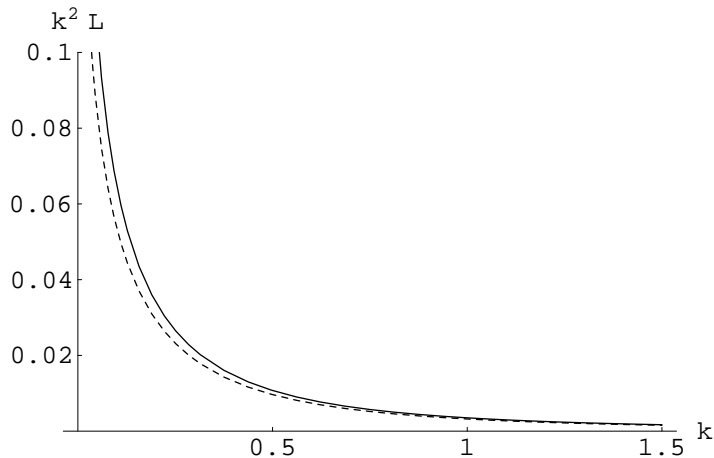}}
\caption{This plot shows the exact one-loop scalar QED effective
Lagrangian (\protect{\ref{1lsclag}}) [solid curve] in comparison with the
leading Borel approximation in  (\protect{\ref{1lscborelleading}})
[dashed curve]. We have multiplied through each function by a common
factor of $(4\pi)^2\kappa^2/m^4$ for ease of presentation. Note that the
agreement is much better than for the weak-field (large $\kappa$)
expansion shown in Figs. 1 and 2.}
\label{f5}
\end{figure}

To begin, we write the one-loop weak-field expansion coefficients
$c_n^{(1)}$ in (\ref{cn1}) in a form which we can easily relate to
(\ref{general}). This is clearly already true of the leading behavior
(\ref{1lscgrowth}), but we can also quantify the corrections. Recall 
Euler's formula (\ref{bernoulli}) that relates the Bernoulli numbers to the Riemann
zeta function.
We shift the index of the expansion coefficients so that the
weak-field summation in (\ref{1lscweak}) begins at $n=0$, as in
(\ref{exp}). Thus, we define new expansion coefficients
\begin{eqnarray}
a_n^{(1)}\equiv c_{n+2}^{(1)}&=& -\frac{{\cal B}_{2n+4}}{(2n+2)(2n+4)}\nonumber\\
&=& 2\frac{(-1)^n}{(2\pi)^{2n+4}}
\left[\Gamma(2n+3)+\Gamma(2n+2)\right]\, \left\{1+\frac{1}{2^{2n+4}}+
\frac{1}{3^{2n+4}}+\dots\right\}
\label{an1}
\end{eqnarray}
Keeping the exponentially leading term 1 in the curly bracketed
factor, we see that the coefficient $a_n^{(1)}$ is the sum of two terms,
each of which has the form in (\ref{general}), with $\mu=2$,
$\rho=1/(2\pi)^2$, and $\nu=3$ and $2$ respectively. So, a direct
application of the Borel integral expression (\ref{genborel}) leads to the
leading Borel approximation
\begin{eqnarray}
{\cal L}^{(1)}_{\rm leading}(\kappa) =\frac{m^4}{(4\pi)^2}\frac{1}{2\pi^4
\kappa^2}\,\int_0^\infty dt\, \,
\frac{e^{-2\kappa t}}{1+\frac{t^2}{\pi^2}} (t+2\kappa\,
t^2)
\label{1lscborelleading}
\end{eqnarray}
Figure \ref{f5} shows a comparison of this leading Borel approximation
with the exact effective Lagrangian (as before, each with an extra
factor of
$\kappa^2$). Note that it does very well at large $\kappa$ (weak field),
but less well at small $\kappa$ (strong field). Nevertheless, it does much
better in the strong field regime than the partial sums of the weak-field
expansion (compare with Fig. \ref{f2}). 

Successive subleading corrections to this leading Borel approximation
follow from considering the (exponential) corrections inside the curly
brackets in the expression (\ref{an1}) for the expansion coefficients.
Once again, these terms are of the "standard form" in (\ref{general}),
now with $\rho=1/(2\pi k)^2$, for $k=2,3,\dots$. Thus, each term can be resummed
using the Borel formula (\ref{genborel}). This leads to
\begin{eqnarray}
{\cal L}^{(1)}_{\rm borel}(\kappa) =\frac{m^4}{(4\pi)^2}\frac{1}{2\pi^4
\kappa^2}\sum_{k=1}^\infty\frac{1}{k^4}\, \int_0^\infty dt\, \,
\frac{e^{-2\kappa t}}{1+\frac{t^2}{k^2\pi^2}} (t+2\kappa\,
t^2)
\label{1lscborel}
\end{eqnarray}
Keeping successive terms in the expansion (\ref{1lscborel}) rapidly improves 
the agreement beyond that shown in Figure \ref{f5} for the leading Borel
approximation (\ref{1lscborelleading}).

In fact, the expression (\ref{1lscborel}) could have been derived directly
by interchanging the summation and integration. To see this, note that
by an integration by parts we can rewrite the proper-time form
(\ref{1lsclag}) of the effective Lagrangian as
\begin{eqnarray}
{\cal
L}^{(1)}(\kappa)=\frac{m^4}{(4\pi)^2}\frac{1}{4\kappa^2}\int_0^\infty
dt\,  e^{-2\kappa t}\left(-{\rm coth}\,
t+\frac{1}{t}+\frac{t}{3}\right)\left(\frac{1}{t^2}+
\frac{2\kappa}{t}\right)
\label{byparts}
\end{eqnarray}
In the large $\kappa$ limit we can use the small $t$ expansion of the coth 
function \cite{abramowitz}
\begin{eqnarray}
{\rm coth}\,t-\frac{1}{t}-\frac{t}{3}=-\frac{2}{\pi^4}\sum_{k=1}^\infty
\frac{t^3}{k^4(1+\frac{t^2}{k^2\pi^2})}
\label{identity}
\end{eqnarray}
from which the Borel result (\ref{1lscborel}) follows. Thus, the proper-time
integral representation (\ref{1lsclag}) is precisely the Borel integral
representation of the divergent asymptotic weak-field series (\ref{1lscweak}). And,
conversely, the weak-field series expansion (\ref{1lscweak}) is just the
large $\kappa$ asymptotic expansion of the proper-time integral expression
(\ref{1lsclag}). This is completely analogous to what happens in the case
of a constant magnetic field background
\cite{dh}.

Now we turn to the Borel analysis of the imaginary part of the effective Lagrangian.
Recall that in the magnetic/electric case, the transition from a magnetic to an
electric background involved the perturbative replacement 
$B^2\to -E^2$, or $B\to i E$. 
This is because the Lorentz invariant combination is $(B^2-E^2)$. Similarly, in
the self-dual case, we change from the self-dual 'magnetic' case to the self-dual
'electric' case by the the analytic continuation
$\kappa\to i\kappa$. Under this transformation, the only change in the
weak-field expansion (\ref{1lscweak}) is that it now becomes a {\it non-alternating}
divergent series. Thus, according to the Borel dispersion relation formula
(\ref{genimag}), the effective Lagrangian acquires an imaginary part. Using this
Borel  dispersion relation formula (\ref{genimag}), together with the explicit
expression (\ref{an1}) for the expansion coefficients (but now with the alternating
$(-1)^n$ factor suppressed), we find
\begin{eqnarray}
{\rm Im} \left[{\cal
L}^{(1)}(i\kappa)\right]
=\frac{m^4}{(4\pi)^3}\frac{1}{\kappa^2}\sum_{k=1}^\infty
\left(\frac{2\pi\kappa}{k}+\frac{1}{k^2}\right)\, e^{-2\pi k \kappa}
\label{1lscimag}
\end{eqnarray}
This agrees precisely with what one finds by a direct analytic
continuation of the proper-time integral representation (\ref{1lsclag}),
together with a principal parts prescription for the poles.

\subsection{Borel Analysis of Two-Loop Case (Scalar QED)} 

The Borel analysis at two-loop is more complicated because the
corrections to the leading behavior (\ref{2lscgrowth}) of the
two-loop weak-field expansion coefficients (\ref{cn2}) are considerably
more complicated than for the one-loop case (\ref{an1}) where the
corrections were simple exponentials. Nevertheless, an analysis of the
large order behavior of the two-loop coefficients (\ref{cn2}) uncovers a
remarkably rich number theoretic structure, and permits a Borel analysis
of the complete prefactor to the leading exponential behavior. This will
be shown in this section.

As before, we shift the summation index in the weak-field expansion
(\ref{2lscweak}) so that it begins at $n=0$. So we define the expansion
coefficients 
\begin{eqnarray}
a_n^{(2)}\equiv c_{n+2}^{(2)}
\label{an2shift}
\end{eqnarray}
where $c_n^{(2)}$ are the two-loop weak-field expansion coefficients given
by (\ref{cn2}). In the Appendix we show that if we neglect sub-leading
exponential corrections (this is analogous to just taking the 1 term in
the curly brackets in (\ref{an1}) in the one-loop case),
then\footnote{The derivation of this expansion requires some nontrivial
results in number theory. This is explained in the appendix.} 
as $n\to\infty$,
\begin{eqnarray}
a_n^{(2)}\sim 2\frac{(-1)^n}{(2\pi)^{2n+4}}\left[
\Gamma(2n+3)-\frac{3}{4}\Gamma(2n+2) -\frac{3}{(2\pi)^2}\sum_{l=2}^{n+1}
\frac{(-1)^l (2\pi)^{2l} {\cal B}_{2l}}{2l}\Gamma(2n+4-2l)\right]
\non\\
\label{an2mod}
\end{eqnarray}
This should be compared with the analogous formula for the one-loop case
when we make the same approximation of ignoring the exponential
corrections in (\ref{an1}):
\begin{eqnarray}
a_n^{(1)}\sim 2\frac{(-1)^n}{(2\pi)^{2n+4}}
\left[\Gamma(2n+3)+\Gamma(2n+2)\right]
\label{an1mod}
\end{eqnarray}
In each case we have written the expansion coefficients as a sum of
gamma functions, each term of which has the general form in
(\ref{general}). The difference is that at one-loop there are only two
terms, while at two-loop there is an infinite series of terms. This
translates directly into the fact that in the one-loop imaginary case,
the prefactor of the exponential in (\ref{1lscimag}) has two terms,
while in the two-loop case we will find an infinite series of terms in
the prefactor (see (\ref{2lscimag}) below).

Given the expansion (\ref{an2mod}) it is straightforward to use the
Borel formula (\ref{genborel}) to write the leading Borel approximate
integral representation for the two-loop effective Lagrangian:
\begin{eqnarray}
{\cal L}^{(2)}_{\rm leading}(\kappa)=\alpha \pi \frac{m^4}{(4\pi)^2}
\frac{1}{\pi^4\kappa} \int_0^\infty dt\,\frac{e^{-2\kappa t}
t^2}{1+\frac{t^2}{\pi^2}}\left[1-\frac{3}{8\kappa t}-
\frac{6\kappa t}{(2\pi)^2} \sum_{l=2}^\infty
\frac{(-1)^l{\cal B}_{2l}}{2l(\frac{\kappa t}{\pi})^{2l}}\right]
\label{2lscborel}
\end{eqnarray}
Using the expansion \cite{abramowitz}:
\begin{eqnarray}
Re \left[\psi(1+iy)\right]=\ln y+\sum_{l=1}^\infty \frac{(-1)^{l-1}{\cal
B}_{2l}}{2l y^{2l}}
\label{realpsi}
\end{eqnarray}
we can rewrite the leading Borel approximation (\ref{2lscborel}) as
\begin{eqnarray}
{\cal L}^{(2)}_{\rm leading}(\kappa)=\alpha \pi \frac{m^4}{(4\pi)^2}
\frac{1}{\pi^4\kappa} \int_0^\infty dt\,\,\frac{e^{-2\kappa t}
t^2}{1+\frac{t^2}{\pi^2}}\left[1-\frac{1}{2\kappa t}-
\frac{6\kappa t}{(2\pi)^2} \left(Re
\left(\psi(1+\frac{i\kappa t}{\pi})\right)-\ln\left(\frac{\kappa
t}{\pi}\right)\right)\right]
\non\\
\label{2lscborel2}
\end{eqnarray}

\begin{figure}[ht]
\centerline{\includegraphics[scale=1]{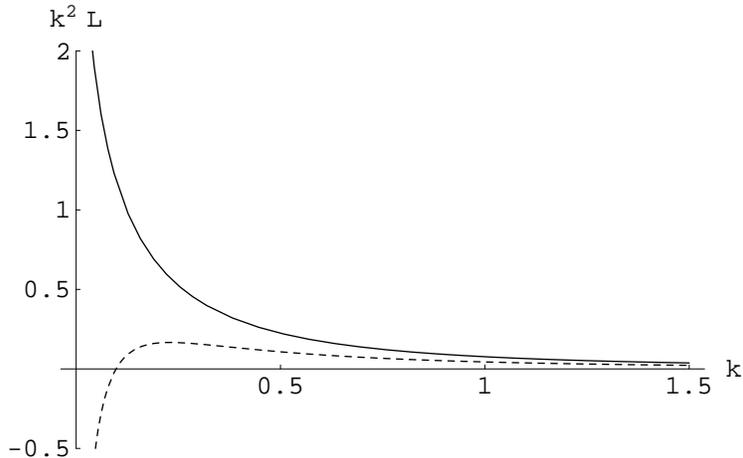}}
\caption{This plot shows the exact two-loop scalar QED effective
Lagrangian (\protect{\ref{2lscintro}}) [solid curve] in comparison with the
leading Borel approximation in  (\protect{\ref{2lscborel2}})
[dashed curve]. We have multiplied through each function by a common
factor of $(4\pi)^2\kappa^2/m^4$ for ease of presentation. Note that the
agreement is not as good in the small $\kappa$ (i.e., strong field) region
as that obtained in the one-loop case in Fig. \protect{\ref{f5}}. See text for
an explanation.}
\label{f6}
\end{figure}

In Figure \ref{f6} we compare this leading Borel integral representation
(\ref{2lscborel2}) with the exact two-loop effective Lagrangian
(\ref{2lscintro}). Notice that the agreement is not as good as in the one-loop
case (compare with Figure \ref{f5}). This can be traced to the fact that the
second term in the expansion (\ref{an2mod}) of the two-loop weak-field 
expansion coefficients is negative. On the other hand, the
second term in the expansion (\ref{an1mod}) of the one-loop weak-field 
expansion coefficients is positive. This second term, in each case, leads to a
log divergence as $\kappa\to 0$. In the one-loop case  this log divergence has
the same sign as the log divergence of the full answer (see
(\ref{1lscstrongleading})), but in the two-loop 
case the log divergence coming from the leading Borel expression 
(\ref{2lscborel2}) has the opposite sign 
compared to the log divergence of the exact answer in the strong field 
(small $\kappa$) limit - compare with (\ref{2lscstrongleading}). Hence, this
mis-match becomes more dramatic at small $\kappa$, as is clearly seen in 
Figure \ref{f6}.

Now consider the analytic continuation $\kappa\to i \kappa$. The only
difference in the weak-field expansion is that the series
(\ref{2lscweak}) becomes a non-alternating series. Thus, according to
the general Borel dispersion relation, there should be a series of
imaginary parts, which can be deduced from the formula (\ref{genimag}).
This leads immediately to (here we use a principal parts prescription for
the pole at $s=\pi$) 
\begin{eqnarray}
{\rm Im}\left[{\cal L}^{(2)}_{\rm leading}(i\kappa)\right]=\alpha \pi
\frac{m^4}{(4\pi)^2}
\frac{1}{2\kappa}
\left[1-\frac{1}{2\pi\kappa}-
\frac{3\kappa}{2\pi} \sum_{l=1}^\infty
\frac{(-1)^l{\cal B}_{2l}}{2l{\kappa}^{2l}}\right]\, e^{-2\pi\kappa}
\label{2lscimag}
\end{eqnarray}
Note that the dominant part, at large $\kappa$, of the prefactor (the $1$ in the
square brackets) is exactly the same as the one-loop leading term in
(\ref{1lscimag}), except for an overall factor of $\alpha \pi$. In the language of
the Borel analysis, this can be traced directly to the fact that the leading growth
rates of the one-loop and two-loop expansion coefficients in (\ref{1lscgrowth}) and
(\ref{2lscgrowth}), respectively, agree precisely. (The overall factor of
$\alpha\pi$ in the two-loop case was explicitly separated out in front of the
two-loop weak-field expansion (\ref{2lscweak})). This is the same behavior as is
found in the case of a constant electric field background, where the leading two-loop
imaginary part is $\alpha \pi$ times the leading one-loop imaginary part [compare
Equations (\ref{L1scalim}), (\ref{fullimag2loop}) and (\ref{expK})], and which is
reflected also in the same leading growth rates of the one- and two-loop weak-field
expansion coefficients \cite{ds1}.

We can now compare this with the imaginary part obtained from the
exact closed-form expression (\ref{2lscintro}). Using the fact that
\cite{abramowitz}
\begin{eqnarray}
{\rm Im}\left[\psi(i\kappa)\right]=\frac{1}{2\kappa}+\frac{\pi}{2}{\rm
coth}(\pi\kappa)
\end{eqnarray}
we deduce that
\begin{eqnarray}
{\rm Im}\left[{\cal L}^{(2)}(i\kappa)\right]&=&\alpha \pi
\frac{m^4}{(4\pi)^3}\frac{1}{2\kappa^2}\left[3\kappa^2
Re\left(\tilde{\psi}(i\kappa)\right)
-1-\kappa\frac{d}{d\kappa}\right]({\rm coth}(\pi\kappa)-1)\nonumber\\
&=&\alpha \pi
\frac{m^4}{(4\pi)^3}\frac{1}{\kappa^2}\sum_{k=1}^\infty \left[3\kappa^2
Re\left(\tilde{\psi}(i\kappa)\right)
-1+2\pi\kappa k\right]\, e^{-2\pi\kappa k}\nonumber\\
&=& \alpha \pi
\frac{m^4}{(4\pi)^2}
\frac{1}{2\kappa}\sum_{k=1}^\infty 
\left[k-\frac{1}{2\pi\kappa}-
\frac{3\kappa}{2\pi} \sum_{l=1}^\infty
\frac{(-1)^l{\cal B}_{2l}}{2l{\kappa}^{2l}}\right]\, e^{-2\pi\kappa k}
\label{2lscimagexact}
\end{eqnarray}
where we have defined the shorthand function
\begin{eqnarray}
\tilde{\psi}(x)\equiv \psi(x)-\ln x+\frac{1}{2x}
\label{psitilde}
\end{eqnarray}
Thus, the leading Borel expression (\ref{2lscimag}) is simply the $k=1$
term (i.e., the leading exponential contribution) in this
expansion. This is completely consistent with our earlier approximation
of only keeping power-law (but not exponential) corrections to the
expression for the expansion coefficients. So we conclude that the direct
use of the Borel dispersion relation (\ref{genimag}), together with the
implied use of a principal parts prescription, is consistent in the
two-loop case also.

It is interesting to see that in (\ref{2lscimagexact}) we have the
complete Lebedev-Ritus prefactor expansion, for any instanton index $k$.
This should be compared with the two-loop electric field case (\ref{fullimag2loop})
and (\ref{expK}), where only the first few terms of this prefactor are known. In
particular, it is now clear that this expansion is itself a divergent asymptotic
series, a fact which was left undecided by the Lebedev-Ritus analysis. Another
interesting issue is the following: comparing the self-dual case result
(\ref{2lscimagexact}) with the corresponding one-loop result (\ref{1lscimag}) we see
that, for any non-zero `field strength' $f$, the $k$-th prefactor at two-loops will
dominate over the corresponding one-loop term for  sufficiently large $k$. On the
other hand, in the self-dual case, it is straightforward to verify that the {\it
total} two-loop contribution (\ref{2lscimagexact}), when one sums over all $k$, is
always smaller (essentially by a factor of $\alpha\pi$) than the {\it total}
one-loop contribution (\ref{1lscimag}). It would be very interesting to see if a
similar thing happens in the electric background case, where the individual $k$
terms in the sum have a direct physical meaning in terms of a coherent multi-pair
production process
\cite{nikishov}.

\section{Conclusions}
\label{conclusions}
\renewcommand{\theequation}{4.\arabic{equation}}
\setcounter{equation}{0}

To conclude, we have presented an analysis of the imaginary part of the
one-loop and two-loop effective Lagangians for scalar QED in a constant self-dual
background. At both one- and two-loop, these self-dual effective Lagrangians have
many properties in common with the effective Lagrangians for a constant magnetic or
electric field background. They have similar weak- and strong-field expansions. Also,
the correspondence between one- and two-loops is very similar in the
magnetic/electric and self-dual backgrounds. In particular, in the self-dual case we
found that the imaginary part of the two-loop effective Lagrangian has a leading term
equal to the leading term of the imaginary part of the one-loop effective
Lagrangian, multiplied by an overall factor of $\alpha\pi$. In terms of the
weak-field expansion coefficients, this is reflected in the fact that the one-loop
and two-loop expansion coefficients (with a factor of $\alpha\pi$ extracted in the
two-loop case) have identical leading growth rates at large order of perturbation
theory. We stress that this agreement between leading growth rates is highly
nontrivial -- it only occurs if one uses the consistently renormalized mass
parameter in the expansion, and thus it is sensitive to the finite part of the mass
renormalization that enters this analysis at two-loops. These facts also hold for the
constant electric field background. It is also consistent with the exponentiation
factor, $\exp(\alpha\pi)$, found by Affleck {\it et al} for the case of pair
production in a weak electric field at strong coupling in scalar QED \cite{affleck}.
Indeed, based on the expectation that this exponentiation occurs also in the
self-dual case, we conjecture that at loop order $l$, the weak-field expansion
will take the form 
\begin{eqnarray}
{\cal L}^{(l)(SD)}(\kappa)=\frac{(\alpha \pi)^{l-1}}{(l-1)!}\,
\frac{m^4}{(4\pi)^2}\sum_{n=2}^\infty c_n^{(l)}\, \frac{1}{\kappa^{2n}}
\label{conj}
\end{eqnarray}
where for each $l$, the expansion coefficients $c_n^{(l)}$ have the same {\it
leading} large $n$ growth rate as $c_n^{(1)}$. Thus, for scalar QED, the
$c_n^{(l)}$ should grow factorially exactly as in (\ref{1lscgrowth}), while for
spinor QED the {\it leading} large $n$ growth should have an additional factor of
$-2$. This also has implications for the leading large $n$ behavior of the zero
momentum limit of the all "+" helicity amplitudes, as is explained in \cite{sd1}
at one- and two-loop. In particular, this conjecture (\ref{conj}) for the
growth with loop order can be immediately extended to the low energy
limit of the `all +' N-photon amplitude at large N. At first sight this
might seem a puzzling result, since it suggests that the N-photon
amplitude, if only in one helicity component and in the low energy limit,
can be expressed as a {\it convergent} power series in $\alpha$ for
sufficiently large $N$. This runs counter to  expectations based on
well-known general arguments which indicate that the loop expansion
should  yield an aymptotic series. This apparent contradiction might be
resolved either by the appearance of poles in the complex $\alpha$ plane
at higher loop orders, invalidating the naive use of the Borel dispersion
relations, or, what seems much more likely, by a slowing down of the
convergence in N of the ratio ${\Gamma^{(l)(EH)}\over\Gamma^{(1)(EH)}}$
with increasing $l$. Such a non-uniformity of the convergence might be
visible already at the three-loop level, which lends additional
motivation to pursuing the calculations presented here to even higher loop
order. 

We also believe that the results we obtain here for {\it
subleading} behavior, in the self-dual background, can be taken as illustrative
of the constant magnetic/electric case also.  The advantage, however, of the
self-dual case is that there exist closed-form expressions for the two-loop
renormalized effective Lagrangians. This means that the analysis is much more
complete in this case. We were able to find the full imaginary part in the
self-dual 'electric' case, and we showed that this is consistent with what one
obtains by a Borel dispersion relation treatment of the weak-field expansion,
which is a non-alternating divergent series.  At two-loop, the prefactor of each
exponential term in the imaginary part, has itself an asymptotic expansion.  
\vskip 1cm

\noindent
{\bf Acknowledgements:} 
We thank Todd Brun for showing us the identity (\ref{todd}), and
Richard Stanley for informing us about Miki's identity
(\ref{mikiidentity}). Discussions and correspondence with Vladimir  Ritus
are also gratefully acknowledged.  Both authors thank Dirk Kreimer, and
the Center for Mathematical Physics at Boston University, for hospitality
during the final stage of this work. We gratefully acknowledge the
support of the NSF and CONACyT through  a US-Mexico collaborative
research grant, NSF-INT-0122615.

\begin{appendix}

\section{Asymptotic expansion of 
two-loop coefficients in scalar QED}
\label{asymp}

\renewcommand{\theequation}{A.\arabic{equation}}
\setcounter{equation}{0}

In this Appendix we compute the large $n$ - expansion of the
coefficients $c_n^{(2)}$, appearing in the two-loop weak field
expansion for scalar QED, up to subexponential terms. We
recall from (\ref{2lscweak}) that the expansion of 
the final two-loop formula (\ref{2lscintro}) yields the
following expression for the coefficients:
\begin{eqnarray}
c^{(2)}_n =
{1\over (2\pi)^2}\biggl\lbrace
\frac{2n-3}{2n-2}\,{\cal B}_{2n-2}
+\frac{3}{2}\sum_{k=1}^{n-1}
{{\cal B}_{2k}\over 2k}
{{\cal B}_{2n-2k}\over (2n-2k)}
\biggr\rbrace
\label{cn2app}
\end{eqnarray}
An alternative form for these expansion coefficients arises when one writes ${\cal
L}_{\rm scal}^{(1)(SD)}(\kappa)$ as a single proper-time integral (see
Equation (3.17) in part I \cite{sd1}):
\bear
c^{(2)}_n &=&{1\over (2\pi)^2}\biggl\lbrace
\frac{2n-3}{2n-2}\,{\cal B}_{2n-2}
+3\Bigl[\psi(2n+1)-{2\over 2n-1}+\gamma-1\Bigr]
{{\cal B}_{2n}\over 2n}
\non\\
&&\hspace{60pt}
+3\sum_{k=1}^{n-1}
{{2n-2}\choose{2k-2}}
{{\cal B}_{2k}\over 2k}
{{\cal B}_{2n-2k}\over 2n-2k}
\biggr\rbrace
\label{cn2alt}
\ear
This second form is more convenient for studying the large order behaviour of
$c_n^{(2)}$. The equivalence of the formulas (\ref{cn2app}) and (\ref{cn2alt}) is not
immediately obvious. However, it can be shown using Euler's identity for Bernoulli
numbers
\footnote{Identities for sums of products of Bernoulli numbers appear in many
contexts in physics and mathematics \cite{fabpan,huahua}.}
\bear
\sum_{k=1}^{n-1}{2n\choose 2k}{\cal B}_{2k}{\cal B}_{2n-2k}
&=&-(2n+1){\cal B}_{2n}
\label{eulerformula}
\ear
together with a more involved identity known as Miki's identity \cite{miki,gessel}:
\bear
\sum_{k=1}^{n-1}{{\cal B}_{2k}{\cal B}_{2n-2k}\over
  (2k)(2n-2k)}=\sum_{k=1}^{n-1}{{\cal B}_{2k}{\cal B}_{2n-2k}\over
  (2k)(2n-2k)}{2n\choose 2k} +{{\cal B}_{2n}\over n}\left(\psi(2n+1)+
\gamma\right)
\label{mikiidentity}
\ear
It is the second form, in (\ref{cn2alt}), for $c_n^{(2)}$ which we will use in the
following for  obtaining the large $n$ behavior of these coefficients.
To start with, we note that the large $n$ expansion of the first term in the curly
brackets is straightforward, using Euler's formula (\ref{bernoulli}) which
relates the Bernoulli numbers to the zeta function, which is of order 1,
with exponentially small corrections. The difficulty is in the last
term in (\ref{cn2alt}), because the index on each of the Bernoulli
numbers need not be large just because $n$ is large. Thus this last
term, which involves the finite sum of a product of Bernoulli numbers
must be modified somehow to bring it into a more manageable form. 
We therefore use the Euler identity (\ref{eulerformula}) to
rewrite (for $n\geq 2$)
\bear
\sum_{k=1}^{n-1}{{2n-2}\choose{2k-2}}{{\cal B}_{2k}\over 2k}
{{\cal B}_{2n-2k}\over 2n-2k}
&=&-{1\over 2n(2n-1)}\sum_{k=1}^{n-1}
{2n\choose 2k}{\cal B}_{2k}{\cal B}_{2n-2k}\nonumber\\
&&+ 2(-1)^n{(2n-1)!\over(2\pi)^{2n}}\sum_{k=1}^{n-1}
{\zeta (2k)\zeta (2n-2k)\over n-k}\nonumber\\
&=&\frac{{\cal B}_{2n}(2n+1)}{2n(2n-1)}+ 
2(-1)^n{(2n-1)!\over(2\pi)^{2n}}\sum_{k=1}^{n-1}
{\zeta (2k)\zeta (2n-2k)\over k}\non\\
\label{reexpressthird}
\ear
The advantage of these manipulations is that we can now separate out the
leading large $n$ growth rate of the coefficients and write:
\bear
c^{(2)}_n
&=&
(-1)^n{(2n-1)!\over (2\pi)^{2n}}
\biggl\lbrace
{n-{3\over 2}\over (n-1)(n-\half)}
\zeta(2n-2)
-{6\over(2\pi)^2}\Bigl[
\psi(2n+1)+\gamma\Bigr]
\,\zeta(2n)
\non\\
&&\hspace{70pt}
+{6\over(2\pi)^2}\sum_{k=1}^{n-1}
{1\over k}
\zeta(2k)\zeta(2n-2k)
\biggr\rbrace
\label{c2zeta}
\ear
In the approximation which neglects exponentially suppressed
terms, the large $n$ behavior of the first two terms is simple, since
the $\zeta$ - factors can be replaced by unity. The large order behavior
of the $\psi$ function follows from using $\psi(m) = \psi(m-1)+{1\over
m-1}$, and the asymptotic expansion(\ref{psiasymptotic}).
Thus, we find (here and in the following, $\sim$ denotes equality
up to terms which are exponentially small for
large $n$)
\bear
{n-{3\over 2}\over (n-1)(n-\half)}
\zeta(2n-2)
&\sim&
{1\over n-1} + \sum_{m=2}^{\infty}
{(-1)^{m+1}
\over 2^{m-2}(n-1)^m}
\non\\
\Bigl[
\psi(2n+1)+\gamma\Bigr]
\,\zeta(2n)
&\sim &
\log  (n-1) + \log(2) +\gamma 
+{5\over 4}{1\over n-1}
\non\\&&
+\sum_{m=2}^{\infty}
{1\over (n-1)^m}
\Bigl[
(-1)^{m+1}(\half + {1\over 2^m})
-{{\cal B}_m\over 2^m m}
\Bigr]
\label{easytermsasymp}
\ear
Note that it will turn out to be convenient to perform the
expansion in $n-1$ rather than in $n$.

The third term in (\ref{c2zeta}) is the problematic one. It involves a
folded sum of zeta functions. While in many cases such sums can be
reexpressed in a form which involves the zeta function only linearly
\cite{fabpan,huahua}, for the case at hand no such
formula appears to be known. However, we can still 
obtain its asymptotic expansion from Euler's formula
(\ref{eulerformula}), in the following way. Define
\bear
Z_n &\equiv&
\sum_{k=1}^{n-1}
{\zeta(2k)\zeta(2n-2k)\over k}
=
\sum_{r,s=1}^{\infty}
Z_n^{rs}
\non\\
Z_n^{rs} &=&
\sum_{k=1}^{n-1}
{1\over k}
r^{-2k}
s^{2k-2n}
\label{rewritefoldedsum}
\ear
We can immediately discard all terms where not either
$r=1$ or $s=1$, since they are exponentially small.
Thus we have to consider the following three
contributions:

$1.$ $r=s=1$: 
\bear 
Z_n^{11} &=& \sum_{k=1}^{n-1}
{1\over k} = \psi(n) +\gamma
\label{Z11}
\ear
$2.$ $s=1<r$: 
\bear
\sum_{r=2}^{\infty}Z_n^{r1}
&=&
\sum_{r=2}^{\infty}\sum_{k=1}^{n-1}
{1\over kr^{2k}}
\sim
\sum_{r=2}^{\infty}\sum_{k=1}^{\infty}
{1\over kr^{2k}}
=
-\sum_{r=2}^{\infty}
\log \Bigl(1-{1\over r^2}\Bigr)
= \log (2)
\label{Zr1}
\ear
$3.$ $r=1<s$:
\bear
\sum_{s=2}^{\infty}Z_n^{1s}
&=&
\sum_{s=2}^{\infty}\sum_{k=1}^{n-1}
{1\over ks^{2n-2k}}
=
\sum_{s=2}^{\infty}\sum_{k=0}^{n-2}
{1\over (n-1-k)s^{2k+2}}
\non\\
&=&
\sum_{m=0}^{\infty}
{1\over (n-1)^{m+1}}
\sum_{s=2}^{\infty}
\sum_{k=0}^{n-2}
{k^m\over s^{2k+2}}
\sim
\sum_{m=0}^{\infty}
{1\over (n-1)^{m+1}}
\sum_{s=2}^{\infty}
\sum_{k=0}^{\infty}
{k^m\over s^{2k+2}}
\non\\
&=&
\sum_{m=0}^{\infty}
{1\over (n-1)^{m+1}}
\sum_{s=2}^{\infty}
{1\over s^2}
\Bigl(
{1\over s^2}
{d\over d ({1\over s^2})}
\Bigr)^m
{1\over 1-{1\over s^2}}
\non\\
&=&
{3\over 4}{1\over n-1}
+
\sum_{m=1}^{\infty}
{1\over (n-1)^{m+1}}
\sum_{k=1}^m {\cal S}^{(k)}_m
k!\sum_{s=2}^{\infty}{1\over (s^2-1)^{k+1}}
\label{Z1s}
\ear

In the last step we used the combinatorial
``normal ordering'' identity
\bear
\Bigl(x{d\over dx}\Bigr)^m
&=&
\sum_{k=1}^m {\cal S}^{(k)}_m x^k\Bigl({d\over dx}\Bigr)^k
\label{todd}
\ear
$(m> 0)$, 
where ${\cal S}^{(k)}_m$ denotes the Stirling number of the second
kind, which is the number of ways of partitioning a set of $m$
elements into $k$ non-empty subsets. Explicitly:
\bear
{\cal S}^{(k)}_m &=&
\sum_{i=0}^k (-1)^i{(k-i)^m\over i!(k-i)!}
\label{stirling}
\ear
The $s$ - sum can be done in closed form after using a partial
fraction decomposition to write
\bear
z_k&\equiv& \sum_{s=2}^{\infty} {1\over (s^2-1)^k} = {(-1)^k\over
2^{2k}} \Biggl\lbrace \sum_{l=1}^{[{k\over 2}]} 2^{2l+1}
{2k-2l-1\choose k-2l} \zeta(2l) -\sum_{l=1}^k (1+2^l) {2k-l-1\choose
k-l} \Biggr\rbrace \non\\ &=& {(-1)^k\over 2^{2k}} \Biggl\lbrace
\sum_{l=1}^{[{k\over 2}]} (-1)^{l+1} {2^{2l}(2\pi)^{2l}\over (2l)!}
{2k-2l-1\choose k-2l} {\cal B}_{2l} -2^{2k-1}\biggl\lbrack 1 + {\Gamma
(k+\frac{1}{2})\over\sqrt{\pi} k!}  \biggr\rbrack \Biggr\rbrace 
\label{computesum}
\ear

Thus, again using (\ref{psiasymptotic})
we obtain the asymptotic expansion of $Z_n$:
\bear
Z_n \sim 
\log (n-1) +\gamma +\log (2)
+{5\over 4}{1\over n-1}
+
\sum_{m=2}^{\infty}
{1\over (n-1)^m}
\biggl[
\sum_{k=1}^{m-1}k!{\cal S}_{m-1}^{(k)}z_{k+1}-{{\cal B}_m\over m}
\biggr]
\label{Znasymp}
\ear
where the $z_k$ are the numbers defined in (\ref{computesum}).
Finally, putting everything together, the final result for the
asymptotic expansion of the coefficients $c_n^{(2)}$ becomes,
up to subexponential corrections,
\bear
c_n^{(2)}
&\sim&
(-1)^n{(2n-1)!\over (2\pi)^{2n}}
\Biggl\lbrace
{1\over n-1}
+
\sum_{m=2}^{\infty}
{1\over (n-1)^m}
\biggl\lbrace
{(-1)^{m+1}\over 2^{m-2}}
\non\\&&
+ {6\over (2\pi)^2} \Bigl[
(-1)^m(\half +{1\over 2^m})
+{{\cal B}_m\over m}({1\over 2^m} -1)
+\sum_{k=1}^{m-1}k!{\cal S}^{(k)}_{m-1}z_{k+1}
\Bigr]
\biggr\rbrace
\Biggr\rbrace
\label{cn2asympapp}
\ear
Observe that the leading $O(\log (n-1))$ and $O((n-1)^0)$ terms cancel out in the
sum. This is a nontrivial self-consistency check. We note that, for the constant
term, the occurrence of this cancelation depends on the value of the finite part of
the mass shift in the mass renormalization \cite{sd1}; it happens only if the
renormalized scalar mass is the physical one. That the leading asymptotic behaviour
depends qualitatively on the finite part of the mass shift term was already observed
in the electric field case
\cite{ds1}, although then only by numerical means.

In order to be able to do the two-loop Borel analysis, we would like to
express these expansion coefficients as a series of $\Gamma$ function
terms. First, rewrite (\ref{cn2asympapp}) in the following form
\bear
c_{n}^{(2)} &\sim &
2{(-1)^n\over (2\pi)^{2n}}
\Bigl(\Gamma(2n-1)+\Gamma(2n-2)\Bigr)
\biggl\lbrace
1+\sum_{m=1}^{\infty}{d_m\over (n-1)^m}
\biggr\rbrace
\label{sn1fact}
\ear
where
\bear
d_m=
{(-1)^{m}\over 2^{m-1}}
+ {6\over (2\pi)^2} \Bigl[
(-1)^{m+1}(\half +{1\over 2^{m+1}})
+{{\cal B}_{m+1}\over {m+1}}({1\over 2^{m+1}} -1)
+\sum_{k=1}^m k!{\cal S}^{(k)}_m z_{k+1}
\Bigr]
\label{dm}
\ear
Next, by standard combinatorics one can derive the
following rearrangements, which are formally
true for any sequence $d_m$:
\bear
1+\sum_{m=1}^{\infty}{d_m\over n^m}
&=&
1+{e_1\over 2n}+{e_2\over 2n(2n-1)}
+{e_3\over 2n(2n-1)(2n-2)} +\ldots,\non\\
e_m &=& \sum_{k=1}^m 2^kS_{m-1}^{(k-1)}d_k
\label{rearre}
\ear
and
\bear
1+\sum_{m=1}^{\infty}{d_m\over n^m}
&=&
1+{f_1\over 2n-1}
+{f_2\over (2n-1)(2n-2)}
+{f_3\over (2n-1)(2n-2)(2n-3)}
+\ldots, \non\\
f_m &=& \sum_{k=1}^m 2^kS_m^{(k)}d_k
\label{rearrf}
\ear
In these formulas, the $S_m^{(k)}$ denote the
Stirling numbers of the {\sl first} kind.
With the help of these formulas, we can rewrite
(\ref{sn1fact}) as
\bear
c_{n}^{(2)} &\sim &
2{(-1)^n\over (2\pi)^{2n}}
\sum_{m=1}^{\infty}
\sigma_{1,m}^{(2)}
\Gamma (2n-m), 
\label{sn1toGamma}
\ear
where
\bear
\sigma_{1,1}^{(2)} &=& 1, \non\\
\sigma_{1,2}^{(2)} &=& e_1 + 1 \non\\
\sigma_{1,m}^{(2)} &=& e_{m-1}+f_{m-2}, \qquad m\ge 3 
\label{sigma}
\ear
Using the explicit form (\ref{dm}) of $d_m$
we have found, using MATHEMATICA,  
the following remarkably complex identity (valid for $m\geq 3$):

\bear
\sum_{k=1}^{m-2} 2^k d_k \left( S^{(k-1)}_{m-2}+S^{(k)}_{m-2}\right)+
2^{m-1}d_{m-1}=\frac{3(2\pi)^{m-2}}{m}\, |{\cal B}_m|
\label{bigid}
\ear
Here $d_k$ are defined in (\ref{dm}). 
This identity has the consequence that the expansion coefficients in the
decomposition (\ref{sn1toGamma}) are finally very simple:
\bear
\sigma_{1,1}^{(2)} &=& 1 \non\\
\sigma_{1,2}^{(2)} &=& -{3\over 4}\non\\
\sigma_{1,m}^{(2)} &=& 3{(2\pi)^{m-2}\over m}
\mid {\cal B}_m \mid,\qquad m\ge 3 
\label{sigmafin}
\ear
In other words, we have shown that, up to exponentially small
corrections, as $n\to\infty$, 
\begin{eqnarray}
c_n^{(2)}&\sim& 2\frac{(-1)^n}{(2\pi)^{2n}}\Gamma(2n-1)\left[ 1-
\frac{3}{4(2n-2)} -\frac{3}{(2\pi)^2}\sum_{l=2}^{n-1}
\frac{(-1)^l (2\pi)^{2l} {\cal B}_{2l}}{2l(2n-2)(2n-3)\dots
(2n-2l)}\right]\nonumber\\
&=&2\frac{(-1)^n}{(2\pi)^{2n}}\left[
\Gamma(2n-1)-\frac{3}{4}\Gamma(2n-2) -\frac{3}{(2\pi)^2}\sum_{l=2}^{n-1}
\frac{(-1)^l (2\pi)^{2l} {\cal B}_{2l}}{2l}\Gamma(2n-2l)\right]
\label{an2modapp}
\end{eqnarray}
This is the expansion needed to perform the Borel analysis of the
two-loop weak field expansion in Section \ref{borelanal} : see
(\ref{an2shift}) and (\ref{an2mod}).

\end{appendix}

\end{document}